\documentstyle[12pt]{article}
\newcommand{\be}{\begin{equation}}
\newcommand{\ee}{\end{equation}}
\newcommand{\bee}{\begin{eqnarray}}
\newcommand{\eee}{\end{eqnarray}}
\begin{document}
\begin{titlepage}
\pagestyle{empty}
\title{Corner transfer matrix of generalised free Fermion vertex systems}
\author{H-P Eckle\thanks{Present address: Department of Physics,
                                          Princeton University,
                                          P.\ O.\ Box 708,
                                          Princeton, New Jersey 08544,
                                          USA}\\
                   \\
        Laboratoire de Physique du Solide,\\
        URA CNRS $n^o 155$, Universit\'e de Nancy I,\\
        BP239, F-54506 Vand\oe uvre l\`es Nancy Cedex, France\\
                                                             \\
      \ and\\
            \\
        T T Truong\\
                   \\
        Laboratoire de Mod\`eles de Physique Math\'ematique\\
        Universit\'e de Tours, Parc de Grandmont, F-37200 Tours, France\\
\mbox{   }\\
\mbox{   }\\
\mbox{   }\\
\mbox{   }
\mbox{   }\\
\mbox{   }\\}
\date{}
\maketitle
\newpage

\begin{abstract}
The Hamiltonian limit of the corner transfer matrix (CTM) of a generalised
free Fermion vertex system of finite size leads to a quantum
spin Hamiltonian of the particular form:
\[
{\cal H}_N=-\sum_{n=1}^{N-1}\left\{ n\left(
				\sigma_n^x\sigma_{n+1}^x
			+\lambda\sigma_n^y\sigma_{n+1}^y
  			+h(\sigma_n^z+\sigma_{n+1}^z)
			           \right)\right\}
\]
Diagonalisation may be achieved for all pairs of parameters
$(\lambda,h)$ with the use of some new elliptic
polynomials which extend the class of special polynomials known so far in the
context of CTM.
\end{abstract}
\end{titlepage}
\pagestyle{plain}
\baselineskip 24pt
\section{Introduction}
Recently we have studied the corner transfer matrix (CTM) of a free Fermion
8-vertex system at criticality \cite{et92}. This investigation has been part of
a
series of studies devoted to special cases in the parameter space of this model
where an explicit solution can be
found for the diagonalisation of such CTM's.
The CTM is an interesting object in the statistical mechanics of
two-dimensional lattice models which was introduced
by Baxter in the seventies (see \cite{baxtersbook} for a review and
references to the original publications), and
with which he was able to compute the mean magnetisation (or polarisation) in
exactly solvable models, such as the eight-vertex model. Baxter has shown then
that in the infinite limit of the two-dimensional lattice the spectrum of the
``generator'' of the CTM is an equidistant spectrum in some domain of the
coupling constants. This result has been largely confirmed by numerous
computations in vertex models of the RSOS type \cite{kyo}.

The pecularity of this spectrum has aroused interest in the last decade for
another discovery in critical phenomena has a similar structure. The
hypothesis of conformal invariance of critcal systems advocated by
Belavin, Polyakov and Zamolodchikov \cite{bpz84}
leads to the conclusion that the generators of row to row transfer
matrices \cite{bcn86} as well as of CTM \cite{ctmgen} of finite-size critical
systems have also equidistant spectra. It is therefore natural to ask whether
there exists a relation between these observations.
In an attempt to clarify this
question we have undertaken a systematic study of the CTM of the simplest
soluble system, the free Fermion system (or its equivalent vertex model).
We shall place ourselves usually at arbitrary temperature $T$ and consider
finite but large systems of size $N$. Therefore both limits $T\rightarrow T_c$
and $N\rightarrow \infty$ can be taken independently.

Let us summarise what has been obtained so far in the study of the CTM of the
free fermion system. The most general free Fermion
system depends on two parameters: the anisotropy parameter $\lambda$, which
measures temperature, and the reduced external magnetic field
$h$ \cite{tp90}. From the
standard method of Lieb, Schultz and Mattis \cite{lsm61}
it is known the CTM of a generalised free Fermion system may be diagonalised by
diagonalising an associated matrix obtained from the eigenvalue equation of
the generator of the CTM. This matix has the peculiar feature that
its eigenfunctions are special polynomials defined by recursion relations.
In reference \cite{tp88} we have solved the simplest case:
\begin{itemize}
\item $\lambda=1$ and $h=0$,
\end{itemize}
which is also a critial line.
The polynomials obtained are particular
cases of Meixner-Pollaczek polynomials. In reference \cite{tp89} we have
considered the case of
\begin{itemize}
\item arbitrary $\lambda$ and $h=0$
\end{itemize}
corresponding to an Ising
model and found two types of Carlitz polynomials of imaginary argument. Then
several other cases are solved in \cite{tp90}:
\begin{itemize}
\item $\lambda=1$ and $h$ arbitrary,
\end{itemize}
where Pollaczek and Gottlieb polynomials are obtained,
\begin{itemize}
\item $\lambda=h^2$, the disorder line \cite{bm71},
\end{itemize}
where Gottlieb polynomials are found. It is remarkable that the
third class of Carlitz polynomials are also found in the CTM of the discrete
Gaussian model \cite{pt91}. Lately we have obtained also the generalised
Pollaczek polynomials in a free Fermion vertex model with a line of defects
\cite{et93}. Up to now the polynomials encountered are all orthogonal
polynomials already known in the mathematical literature. Recently we have
tackled other regimes in the parameter space $(\lambda,h)$ where the
orthogonality property is not known. First we have studied in \cite{et92} the
polynomials associated with the critcal line
\[
\lambda=2h-1
\]
and in this paper
we shall deal with the general case
\[
\lambda>h^2
\]
where both $\lambda$ and $h$
are not restricted to satisfy any equation.
The remaining regime of the parameter space, $\lambda<h^2$, where additional
mathematical complexities arise, will be treated in a forthcoming publication.

As the reader may have noticed the CTM of free Fermion systems is a simple
device of statistical mechanics which introduces the special polynomials.
The spectrum of a CTM of size $N$ is ``essentially'' given by the zeros of a
polynomial of size $N$, when $N$ is sufficiently large. Hence, as one may
guess, the distribution of zeros tends to be a uniform distribution as
$N\rightarrow \infty$. This fact confirms the findings of Baxter and verifies
naturally the predictions of conformal invariance in critical systems.

The new feature in this paper is the appearance of new ``elliptic'' polynomials
which generalise the only known types of elliptic polynomials discovered
by Carlitz \cite{carlitz60}. Some properties of these polynomials including
the asymptotic distribution of their zeros, hence the eigenvalue
spectrum of the system, will be given.

Physically the ``time'' generator of such free Fermion 8-vertex problems is
simply the anisotropic $XY$ quantum spin chain in a magnetic field $h$.
The counterpart
of this using the CTM approach is the following generator for a finite chain of
$N$ sites:
\be
L_0 = \sum_{n=1}^{N-1} \{ n ( \sigma_n^x \sigma_{n+1}^x +
                       \lambda \sigma_n^y \sigma_{n+1}^y) +
			h (2 n - 1) \sigma_n^z \} + h (N-1) \sigma_N^z
\label{eq:gen}
\ee
where $\lambda$ is the anisotropy parameter describing essentially
the temperature. Note the linear increase of the coupling strength along the
chain.
The problem of the simple chain, i.\ e.\ without that specific linear increase
of the couplings, has been fully solved long ago \cite{xy}.
But the generator $L_0$ has not yet, to our knowledge, been explicitly
solved for general values of $\lambda$ and $h$.

The paper is organised as follows. In section 2 we outline the method employed
for calculating analytically the eigenvalues of the generator $L_0$ which is
based on
the introduction of generating functions for the components of the eigenvectors
of $L_0$. The recursion relations derived for the components of the
eigenvectors
are then equivalent to a set of coupled first--order differential equations
which
are solved formally in section 3. In the solution we encounter an elliptic
integral.
We need to appropriately parametrize and invert this elliptic integral which is
done
in section 4. This leads us to expressions for the generating functions
in section 5
which are used in section 6 to derive explicitly the components of the
eigenvectors
as elliptic polynomials from Cauchy's theorem. In section 7 finally we obtain
an
integral representation for the components of the eigenvectors that can be used
to calculate asymptotically for large system size $N$ the eigenvalues of $L_0$.
Section 8 summarizes our findings and gives an outlook on open problems.
In the
Appendix we give some details for certain limiting cases of the analysis of the
main text.

\section{Method for diagonalising $L_0$}
The standard method we adopt here is that of Lieb, Schultz and Mattis
 \cite{lsm61} which consists in rewriting $L_0$ in terms of Fermion operators.
Then
 one is
left with the diagonalisation of two non-commuting matrices $(A-B)$ and $(A+B)$
in the language of Lieb, Schultz and Mattis. If we denote the components of the
eigenvectors ${\psi}$ and ${\phi}$ by ${\psi}=(\psi_1,\ldots,\psi_N)$
and ${\phi}=(\phi_1,\ldots,\phi_N)$ we have two recursion relations coupling
the components $\psi_n$ and $\phi_n$
\bee
(n - 1) \psi_{n-1}+n\lambda\psi_{n+1}-h(2n-1)\psi_n=\varepsilon\phi_n
\label{eq:rr1}\\
\lambda(n - 1) \phi_{n-1}+n\phi_{n+1}-h(2n-1)\phi_n=\varepsilon\psi_n
\label{eq:rr2}
\eee
For the end components $n=N$ of the finite chain we have only
\bee
(N - 1) \psi_{N-1}-h(N-1)\psi_N=\varepsilon\phi_N
\label{eq:rr3}\\
\lambda(N - 1) \phi_{N-1}-h(N-1)\phi_N=\varepsilon\psi_N
\label{eq:rr4}
\eee
Using the recursion relations (\ref{eq:rr1}) and (\ref{eq:rr2}) the equations
for
the end components (\ref{eq:rr3}) and (\ref{eq:rr4}) are equivalent to
\bee
\phi_{N+1}=h\phi_N \label{eq:bc1}\\
\lambda\psi_{N+1}=h\psi_N \label{eq:bc2}
\eee
which resemble periodic boundary conditions. Note that the components
$\phi_{N+1}$ and $\psi_{N+1}$ are only defined through these equations.

The method employed is simple in principle: it consists to finding an
expression for
$\psi_n$ and $\phi_n$ , then on substituting into eqs.\ (\ref{eq:rr3}) and
(\ref{eq:rr4}) one obtains
the eigenvalues $\epsilon$. However, as we are only interested in large
values of $N$ an asymptotic expression will only be necessary.
For the so far known cases \cite{tp90} the $\psi_n$ and $\phi_n$ are
expressible in terms
of known special polynomials. We expect thus that for
general values of $\lambda$ and $h$ they are also polynomials of the elliptic
type which are seen in the case of the doubled Ising model \cite{tp89}.
In special cases where $\lambda$ and $h$ were chosen to satisfy particular
conditions it was previously possible to identify the recursion relations
(\ref{eq:rr1}) and (\ref{eq:rr2}), after
some trivial
transformations, to the recursion relations for some known polynomials.
Here as in another case discussed recently \cite{et93}
we introduce the generating functions as formal power series in a parameter $t$
with the components of the eigenvectors as coefficients
\bee
\psi(t)=\sum_{n=1}^\infty t^{n-1}\psi_n
\label{eq:gf1}\\
\phi(t)=\sum_{n=1}^\infty t^{n-1}\phi_n
\label{eq:gf2}
\eee
The recursion relations (\ref{eq:rr1}) and (\ref{eq:rr2}) are then equivalent
to a set of coupled first--order differential equations for the
generating functions
\bee
(t^2+\lambda-2ht)\psi^\prime+(t-h)\psi=\varepsilon\phi
\label{eq:fo1}\\
(\lambda t^2+1-2ht)\phi^\prime+(\lambda t-h)\phi=\varepsilon\psi
\label{eq:fo2}
\eee
{}From the explicit solutions $\psi(t)$ and $\phi(t)$ of these differential
equations
with the given initial conditions one may then extract $\phi_n$ and $\psi_n$ by
using the Cauchy theorem.
Then for large $N$ the boundary conditions (\ref{eq:bc1}) and (\ref{eq:bc2})
determine the spectrum of $L_0$ as in previous works of this series
\cite{et92,tp90}. Since the presence of two parameters $\lambda$ and $h$
complicates the mathematical working considerably we shall proceed step by step
in presenting the solution. The central problem at hand is simply the
parametrization by elliptic functions of the solution of the coupled set of
differential equations (\ref{eq:fo1}) and (\ref{eq:fo2}).

\section{Formal solution of the differential equations -- generating functions}
Before writing down the formal solution of the differential equations
(\ref{eq:fo1}) and (\ref{eq:fo2}), we note that these equations as well as the
recursion relations for $\psi_n$ and $\phi_n$, eqs. (\ref{eq:rr1}) and
(\ref{eq:rr2}), remain globally invariant under the combined transformations
\[
\lambda\rightarrow\lambda^{-1}, \qquad h\rightarrow\lambda^{-1}h
\qquad \mbox{and} \qquad \epsilon\rightarrow\lambda^{-1}\epsilon
\]
followed by
\[
\psi\rightarrow\phi \qquad \mbox{and} \qquad \phi\rightarrow\psi \qquad
(\psi_n\rightarrow\phi_n \qquad \mbox{and} \qquad \phi_n\rightarrow\psi_n)
\]
This property allows us to restrict our study to the domain $S$ defined by
\[
0<\lambda<1 \qquad \mbox{and} \qquad 0<h<1.
\]
Using this symmetry relation integrating
factors can be found \cite{brianshint} and as in \cite{et92} the general
solutions of the differential equations take the form of Meixner's
generating functions \cite{meixner34} (up to constants)
\bee
\psi(t)\propto f(t)\exp(\varepsilon w(t;\lambda,h))
\label{eq:mg1}\\
\phi(t)\propto g(t)\exp(\varepsilon w(t;\lambda,h))
\label{eq:mg2}
\eee
where
\be
f(t)=(t^2+\lambda-2ht)^{-\frac{1}{2}}, \qquad
g(t)=(\lambda t^2+1-2ht)^{-\frac{1}{2}}
\label{eq:prefn}
\ee
and $w=w(t;\lambda,h)$ is the two--parameter elliptic integral
\be
w(t;\lambda,h)=\int_{t_0}^t dx f(x) g(x)
\label{eq:ei}
\ee
The lower bound of the integral will be chosen as to agree with known results.
Note that for $h=0$, one recovers the case of the doubled Ising model studied
in
\cite{tp89}. Generating functions of the type (\ref{eq:mg1}) and (\ref{eq:mg2})
are called generating functions of the Meixner type by Chihara
\cite{chihara78}.
Thus it appears that (\ref{eq:mg1}) and (\ref{eq:mg2}) represent perhaps the
most general case known so far. The main problem here is the inversion of the
two--parameter
elliptic integral $w(t)$, eq.\ (\ref{eq:ei}), and then the representation of
the
$\psi_n$ and $\phi_n$ as contour integrals.

\section{Parametrization and inversion of the elliptic integral $w(t)$}
An essential step in the explicit solution consists in an appropriate
parametrization of the elliptic integral $w(t)$. We shall follow here the
procedure given by Greenhill \cite{greenhill31}.
Both polynomials
$N(t)\equiv t^2 + \lambda - 2 h t$ and $D(t)\equiv \lambda t^2 + 1 - 2 h t$
appearing in $w(t)$ have the same discriminant $\Delta=h^2-\lambda$. The zeros
of
$N(t)$ in the parameter range $0<\lambda<1$ and $0<h<1$, the region $S$, are
\be
\gamma=h+\sqrt{\Delta}, \qquad \delta=h-\sqrt{\Delta} \nonumber
\ee
whereas $D(t)$ has zeros at
\be
\alpha=\frac{\gamma}{\lambda}, \qquad \beta=\frac{\delta}{\lambda}\nonumber.
\ee

The square $S$
is divided into 2 regions by the disorder
line \cite{tp90} $\lambda=h^2$ (see Fig. 1), separating oscillating from
monotonous behaviour of the correlation functions \cite{bm71}.

In the region $S_1$, $\lambda>h^2$, the zeros are pairwise complex conjugate,
whereas in the region $S_2$, $\lambda<h^2$, the zeros are all real and ordered
according to $-\infty<\delta<\gamma<\beta<\alpha<\infty$.

The inversion of the elliptic integral in the region $S_2$ is more involved and
we shall defer its study to a sequel publication. In the remainder of this
paper
we shall be only concerned with the region $S_1$.

The region $S_1$ is limited by three boundaries which contain known
results \cite{tp90}:
\begin{itemize}
\item $h=0$  The doubled Ising model, viewed as free Fermion eight-vertex
             model \cite{tp89}: $\psi_n$ and $\phi_n$ are Carlitz
                  elliptic polynomials of imaginary arguments.
\end{itemize}
\begin{itemize}
\item $\lambda =1$ Isotropic case in the presence of a magnetic field $h$.
                   The solutions $\psi_n$ are given in terms of
                   Meixner Pollaczek polynomials.
\end{itemize}
\begin{itemize}
\item $\lambda=h^2$ The disorder line where the solutions $\psi_n$
                    are expressed in terms of Gottlieb (Meixner polynomials of
                    the first kind) polynomials.
\end{itemize}
Note that on these three boundaries the three types of polynomials are all
orthogonal polynomials, their recursion relations are always reducible to
tridiagonal form. The $\psi_n$ and $\phi_n$ studied here provide an
interpolation
between the three classes of polynomials and are presumably also orthogonal
polynomials.

The elliptic integral $w(t)$ may be inverted in a standard way: one may give
$t$
as a function of $w$ following the example given in the book of Greenhill
\cite{greenhill31}, \S 70. In the following we outline Greenhill's procedure.

Through the change of variables
\be
y(t)=\frac{N(t)}{D(t)}=\frac{t^2+\lambda-2ht}{\lambda t^2+1-2ht}
\label{eq:subs}
\ee
the elliptic integral is transformed from its Jacobian form $w(t)$
into the Weierstra{\ss} form
\be
w(y)=\frac{1}{\sqrt{\lambda-h^2}}\int_{y_0}^y \frac{dx}
					{\sqrt{4x(y_1-x)(x-y_2)}}
\label{eq:wf}
\ee
where $y_0=y(t_0)$ and $y_1$ and $y_2$ are the maximum and minimum values of
$y(t)$,
respectively, at the points $t_1$ and $t_2$ related by
\bee
h^2(1-y_{1,2})^2=(\lambda-y_{1,2})(1-\lambda y_{1,2}), \qquad y_1&=&y_2^{-1}\\
t^2_{1,2}-\frac{1+\lambda}{h}t_{1,2} + 1=0, \qquad t_1&=&t_2^{-1}
\eee
and
\be
y_2=\frac{1}{2(\lambda-h^2)}\left\{(1+\lambda^2-2h^2)-(1-\lambda)
\sqrt{(1+\lambda)^2-4h^2}\right\}<y_1
\label{eq:y2}
\ee

Then from the standard Weierstra{\ss} form one may solve the inversion problem
in three different ways
\bee
w(y) & = & \sqrt{\frac{y_2}{\lambda-h^2}}\mbox{sn}^{-1}
                                 \left(
                                      \sqrt{\frac{y_1-y}{y_1-y_2}},\kappa
                                 \right) \nonumber\\
     & = & \sqrt{\frac{y_2}{\lambda-h^2}}\mbox{cn}^{-1}
                                 \left(
                                      \sqrt{\frac{y-y_2}{y_1-y_2}},\kappa
                                 \right) \label{eq:jacobi}\\
     & = & \sqrt{\frac{y_2}{\lambda-h^2}}\mbox{dn}^{-1}(\sqrt{y_2 y},\kappa)
                                        \nonumber
\eee
in terms of the Jacobian elliptic functions sn$(v,\kappa)$, cn$(v,\kappa)$ and
dn$(v,\kappa)$ of argument $v$ and modulus $\kappa$. In particular the modulus
$\kappa$ is given by $\kappa^2=1-y_2^2$, thus $y_2$ is the complementary
modulus of $\kappa$.
Conversely for later use one can extract from (\ref{eq:jacobi}) the quantities
\bee
y_1-y&=&(y_1-y_2)\mbox{sn}^2(wq,\kappa) \nonumber\\
y-y_2&=&(y_1-y_2)\mbox{cn}^2(wq,\kappa) \label{eq:use}\\
y&=&\frac{1}{y_2}\mbox{dn}^2(wq,\kappa) \nonumber
\eee
with $q=\sqrt{\frac{\lambda-h^2}{y_2}}$.
To obtain $t$ as a function of $w$, i.\ e.\ to invert the elliptic integral
(\ref{eq:ei}), we use equation (102) of Greenhill \cite{greenhill31}
\bee
y_1-y&=&\frac{(\lambda y_1-1)(t_1-t)^2}{\lambda t^2+1-2 h t}
\label{eq:green1}\\
y-y_2&=&\frac{(1-\lambda y_2)(t-t_2)^2}{\lambda t^2+1-2 h t} \label{eq:green2}
\eee
By dividing out these equations and using the first two of eqs.\ (\ref{eq:use})
as well as $y_1\cdot y_2=1$ and $t_1\cdot t_2=1$, we obtain the following form
for $t=t(w)$
\be
t(w)=\frac{t_1\sqrt{\lambda-y_2}
                   \mbox{cn}(qw,\kappa) +
        t_2\sqrt{y_2(1-\lambda y_2)}
                    \mbox{sn}(qw,\kappa)}
       {\sqrt{\lambda-y_2}
                   \mbox{cn}(qw,\kappa) +
        \sqrt{y_2(1-\lambda y_2)}
                    \mbox{sn}(qw,\kappa)}
\ee
Thus in the case of two parameters $\lambda$ and $h$ the inversion of eq.\
(\ref{eq:ei}) yields a rational function of the Jacobi elliptic functions sn
and cn.
However in order to agree with the known results on the boundary line $h=0$,
where
we have a canonical elliptic integral of Legendre form and hence $t$ is simply
proportional to a Jacobi elliptic sn$(v,k)$ function, as was always the case in
previous studies \cite{tp90} on special lines where one had therefore
only one parameter, we transform to imaginary argument and complementary
modulus
$y_2$ ($\kappa^2+y_2^2=1$) with the transformation formulae (Jacobi's imaginary
transformation)
\be
\mbox{sn}(qw,\kappa)=-i\frac{\mbox{sn}(i qw,\kappa^\prime)}
                                    {\mbox{cn}(i qw,\kappa^\prime)}, \qquad
\mbox{cn}(qw,\kappa)=\frac{1}{\mbox{cn}(i qw,\kappa^\prime)}
\ee
Thereby we obtain
\be
t=\frac{t_1 - i \sqrt{y_2}
                \mbox{sn}(iqw,y_2)}
       {1 -  i t_1 \sqrt{y_2}
                \mbox{sn}(iqw,y_2)}
\label{eq:tsans}
\ee

At $h=0$ one may directly invert the elliptic integral of Jacobian form
\[
w(t)=\int_0^t\frac{dx}{\sqrt{(\lambda x^2+1)(x^2+\lambda)}}
\]
to obtain
\[
t(w)=\sqrt{\lambda}\cdot\frac{\mbox{sn}(w,\lambda^\prime)}
     {\mbox{cn}(w,\lambda^\prime)}
\qquad
\lambda^\prime=\sqrt{1-\lambda}
\]
By the transformation to imaginary argument and complementary modulus this
yields
\be
t=-i\sqrt{\lambda}\mbox{sn}(iw,\lambda)
\label{eq:th0}
\ee
Of course we want that (\ref{eq:tsans}) reduces to (\ref{eq:th0}) in the limit
$h\rightarrow0$. To achieve this we shift the argument in (\ref{eq:tsans}) by
$K^\prime$, the complete elliptic integral of the complimentary modulus, i.\
e.\
the quarter period in the imaginary direction of the Jacobian elliptic
functions:
$w=-u+w_0$ with $qw_0=K^\prime(y_2)$, which transforms the function sn
according to
\[
\mbox{sn}(v+iK^\prime,k)=\frac{1}{k\mbox{sn}(v,k)}
\]
Through this last transformation we arrive at
\be
t=\frac{t_2 - i \sqrt{y_2}
                \mbox{sn}(iqu,y_2)}
       {1 - i t_2 \sqrt{y_2}
                \mbox{sn}(iqu,y_2)}
\label{eq:tshift}
\ee
Basically this result means that we have to choose the lower integration bound
in
(\ref{eq:ei}) in such a way that (\ref{eq:tshift}) holds. For $h\rightarrow0$
(i.\ e.\ $y_2\rightarrow\lambda$ and $t_2\rightarrow0$) now (\ref{eq:tshift})
agrees with (\ref{eq:th0}). From now on we shall use this condition at
$h\rightarrow0$ as reference which will also fix the constants in the
generating
functions (\ref{eq:mg1}) and (\ref{eq:mg2}).

\section{Expressions for the generating functions}
{}From (\ref{eq:green2}) we have for the prefactor of the generating function
$\phi(t)$ to be
\[
\sqrt{\lambda t^2 + 1 - 2ht}=\sqrt{\frac{1 - \lambda y_2}{y -y_2}}(t-t_2)
\]
Using the second of eqs.\ (\ref{eq:use}) and performing both transformations
described above consecutively, i.\ e.\ Jacobi's imaginary transformation and
the transformation according to the shifted argument, we obtain together with
(\ref{eq:tshift}) to evaluate $(t-t_2)$
\[
\frac{1}{\sqrt{\lambda t^2 + 1 - 2ht}}=\sqrt{\frac{1 -  y_2^2}{1 - \lambda
y_2}}
                                   \cdot\frac{1-
it_2\sqrt{y_2}\mbox{sn}(iqu,y_2)}
                                            {(1-t_2^2)\mbox{dn}(iqu,y_2)}
\]
For $h\rightarrow0$ this expression has the correct limit \cite{tp89}
\[
\lim_{h\rightarrow\infty}\frac{1}{\sqrt{\lambda t^2 + 1 - 2ht}}=
                               \frac{1}{\mbox{dn}(iqu,\lambda)}
\]
Hence we ontain the generating function of the $\phi_n$ as
\bee
\phi(t) & = & \frac{e^{w\epsilon}}{\sqrt{\lambda t^2 + 1 - 2ht}}\nonumber\\
        & = & \sqrt{\frac{1 -  y_2^2}{1 - \lambda y_2}}\cdot
              \frac{\exp(q^{-1}\epsilon K^\prime(y_2))}{(1-t_2^2)}\cdot
              \frac{1- it_2\sqrt{y_2}\mbox{sn}(iqu,y_2)}
                   {\mbox{dn}(iqu,y_2)}e^{-\epsilon u}
\eee
Similarily using the first of eqs.\ (\ref{eq:use}) and eq.\ (\ref{eq:green1})
we
obtain after the necessary transformations the generating function for the
other
set of components of the eigenvectors $\psi_n$
\bee
\psi(t) & = & \frac{e^{w\epsilon}}{\sqrt{t^2 + \lambda - 2ht}}\nonumber\\
        & = & \sqrt{\frac{1 -  y_2^2}{(1 - \lambda y_2)}}\cdot
              \frac{\exp(q^{-1}\epsilon K^\prime(y_2))}{(1-t_2^2)}\cdot
              \frac{1 - it_2\sqrt{y_2}\mbox{sn}(iqu,y_2)}
                   {\sqrt{y_2}\mbox{cn}(iqu,y_2)}e^{-\epsilon u}
\eee
To simplify the notation we shall set
\be
{\cal N}= \sqrt{\frac{1 -  y_2^2}{(1 - \lambda y_2)}}
          \exp(q^{-1}\epsilon K^\prime(y_2))
\ee
and
\be
\xi=i\sqrt{y_2}\mbox{sn}(iqu,y_2)
\ee
so that
\be
\phi(t)=\frac{{\cal N}}{(1-t_2^2)}\cdot\frac{1-t_2\xi}{\mbox{dn}(iqu,y_2)}
         e^{iq^{-1}\epsilon(iqu)}, \quad \mbox{and} \quad
\psi(t)=\frac{{\cal
N}}{(1-t_2^2)}\cdot\frac{1-t_2\xi}{\sqrt{y_2}\mbox{cn}(iqu,y_2)}
         e^{iq^{-1}\epsilon(iqu)} \label{eq:gffin}
\ee

\section{The elliptic polynomials $\psi_n$ and $\phi_n$}
{}From the generating functions (\ref{eq:mg1}) and (\ref{eq:mg1}) we may
compute
the coefficients $\psi_n$ and $\phi_n$ which are of course functions of the
eigenvalue $\epsilon$ with the help of Cauchy's formula
\[
\psi_n=\frac{1}{2i\pi}\oint \psi(t) t^{-n-1} dt
\]
Substituting $\psi(t)$ by its expression (\ref{eq:gffin}) and using the
variable
$\xi$ with
\[
t=\frac{t_2-\xi}{1-t_2\xi}, \qquad
dt=-\frac{(1-t_2^2)}{(1-t_2\xi)^2}d\xi
\]
we obtain
\be
\psi_n(x)=-\frac{{\cal N}}{2i\pi}\oint
\frac{e^{ixz}}{\sqrt{y_2}\mbox{cn}(z,y_2)}
                                      \frac{(1-t_2\xi)^n}{(t_2-\xi)^{n+1}} d\xi
\label{eq:fromcauchy}
\ee
where $z=iqu$ and $x=q^{-1}\epsilon$, the scaled eigenvalue.

We can express $\psi_n$ in terms of the usual Carlitz polynomials (cf.\
\cite{tp89})
\be
\frac{e^{ixz}}{\mbox{cn}(z,y_2)}=\sum_{p=0}^\infty
                                 \left(\frac{\xi}{\sqrt{y_2}}\right)^p
                                 \frac{P_p^\star(x)}{p!}
\ee
where
\[
P_p^\star(x)=\left\{ \begin{array}{ll}
                           C_p^\star & \mbox{for $p$ even}\\
                           D_p^\star & \mbox{for $p$ odd}
                     \end{array}
             \right.
\]
are the Carlitz polynomials of even and odd order, respectively, associated
with the
Jacobian elliptic functions cn and dn. Then $\psi_n$ takes the form of a
formally
infinite series in these polynomials
\be
\psi_n(x)=-\frac{{\cal N}}{2i\pi}
          \sum_{p=0}^\infty\frac{P_p^\star(x)}{(\sqrt{y_2})^{p+1}p!}
          \oint\xi^p\frac{(1-t_2\xi)^n}{(t_2-\xi)^{n+1}} d\xi
\label{eq:psiform}
\ee
The condition $t=0$ is equivalent to $\zeta=t_2-\xi=0$. Thus the contour
integral
in (\ref{eq:psiform}) may be directly evaluated in the complex $\zeta$ plane
\[
-\oint(t_2-\zeta)^p\frac{(1-t_2^2+t_2\zeta)^n}{\zeta^{n+1}}d\zeta=2i\pi\sigma_n(p)
\]
where
\be
\sigma_n(p)=\sum_{q+m=p}(-1)^{p-q}\left(
                                   \begin{array}{c} p \\ q
                                   \end{array}
                                  \right)
                                  \left(
                                   \begin{array}{c} n \\ m
                                   \end{array}
                                  \right)
                         (1-t_2^2)^{m}(t_2)^{n-m+q}
\ee
We arrive at
\be
\psi_n(x)={\cal N}\sum_{p=0}^\infty\sigma_n(p)\frac{P_p^\star(x)}
                                                  {(\sqrt{y_2})^{p+1}p!}
\label{eq:psicon}
\ee
However the $\psi_n(x)$ are nevertheless polynomials of finite order $n$ in the
variable $x$. To see this we first evaluate (\ref{eq:psicon}) formally for
$n=0$.
We have
\[
\sigma_0(p)=(t_2)^p
\]
and
\[
\psi_0(x)=\frac{{\cal N}}{\sqrt{y_2}}\sum_{p=0}^\infty
                                       \left(
                                         \frac{t_2}{\sqrt{y_2}}
                                       \right)^p\frac{P_p^\star(x)}{p!}
\]

Defining $z_2$ by
\[
t_2=i\sqrt{y_2}\mbox{sn}(z_2,y_2)
\]
we can resum the series and find
\be
\psi_0(x)=\frac{{\cal N}}{\sqrt{y_2}}\frac{e^{ixz_2}}{\mbox{cn}(z_2,y_2)}
\label{eq:psi0}
\ee
which is essentially a constant.
{}From $\psi_0$ taken formally at arbitrary values of $z$
we can compute all $\psi_n$ for $n>0$ and thereby indeed show that $\psi_n$
are polynomials of finite order $n$. For $\psi_1$ we obtain e.\ g.\ the
following formula
\[
\psi_1(x)=\left\{ t_2 - (1-t_2^2)\frac{\partial}{\partial t_2}\right\}
                            \psi_0(x)
\]
which is a polynomial in $x$ of first order after dividing out the exponential.

More generally we can show that $\psi_n$ is a polynomial of order $n$ in $x$
explicitly given by
\bee
\psi_n(x)=\left\{ t_2^n \right.   & - & \left.
                        \left(
                        \begin{array}{c} n \\ n-1
                        \end{array}
                        \right) t_2^{n-1}\frac{(1-t_2^2)}{1!}
                                         \frac{\partial}{\partial t_2}
          +
                        \left(
                        \begin{array}{c} n \\ n-2
                        \end{array}
                        \right)t_2^{n-2}\frac{(1-t_2^2)^2}{2!}
                                        \frac{\partial^2}{\partial
t_2^2}+\cdots
         \right.      \nonumber    \\
         \left. \right.
           \cdots  & + &
         \left.
                        \frac{(-1)^q}{q!}
                        \left(
                        \begin{array}{c} n \\ n-q
                        \end{array}
                        \right) t_2^{n-q}
                               (1-t_2^2)^q
                               \frac{\partial^q}{\partial t_2^q}\cdots \right.
\\
         \left. \right.
           \cdots  & + &
         \left.
                        \frac{(-1)^n}{n!}
                        \left(
                        \begin{array}{c} n \\ n
                        \end{array}
                        \right)(1-t_2^2)^n
                               \frac{\partial^n}{\partial t_2^n}
          \right\}\psi_0(x)
\nonumber
\eee

With the same procedure we show that $\phi_n$ are polynomials in $x$ of finite
order
$n$
\be
\phi_n(x)={\cal N}\sum_{p=0}^\infty\sigma_n(p)\frac{Q_p^\star(x)}
                                                  {(\sqrt{y_2})^p p!}
\label{eq:phicon}
\ee
Here
\[
Q_p^\star(x)=\left\{ \begin{array}{ll}
                           D_p^\star & \mbox{for $p$ even}\\
                           C_p^\star & \mbox{for $p$ odd}
                     \end{array}
             \right.
\]
and we have for $n=0$
\be
\phi_0(x)={\cal N}\frac{e^{ixz_2}}{\mbox{dn}(z_2,y_2)}
\ee
from which we again compute all polynomials $\phi_n(x)$ of order $n>0$
explicitly
\bee
\phi_n(x)=\left\{ t_2^n \right.   & - & \left.
                        \left(
                        \begin{array}{c} n \\ n-1
                        \end{array}
                        \right) t_2^{n-1}\frac{(1-t_2^2)}{1!}
                                         \frac{\partial}{\partial t_2}
          +
                        \left(
                        \begin{array}{c} n \\ n-2
                        \end{array}
                        \right)t_2^{n-2}\frac{(1-t_2^2)^2}{2!}
                                        \frac{\partial^2}{\partial
t_2^2}+\cdots
         \right.      \nonumber    \\
         \left. \right.
           \cdots  & + &
         \left.
                        \frac{(-1)^q}{q!}
                        \left(
                        \begin{array}{c} n \\ n-q
                        \end{array}
                        \right) t_2^{n-q}
                               (1-t_2^2)^q
                               \frac{\partial^q}{\partial t_2^q}\cdots \right.
\\
         \left. \right.
           \cdots  & + &
         \left.
                        \frac{(-1)^n}{n!}
                        \left(
                        \begin{array}{c} n \\ n
                        \end{array}
                        \right)(1-t_2^2)^n
                               \frac{\partial^n}{\partial t_2^n}
          \right\}\phi_0(x)
\nonumber
\eee

\section{Real integral representation and asymptotic behaviour}
In the same spirit as in reference \cite{tp89} we derive now the {\em real}
integral
representation for $\psi_n(x)$, using eq.\ (\ref{eq:fromcauchy}) in the complex
$z$--plane. As before in \cite{tp89} we choose the rectangular contour
$\mbox{Im}(z)=\pm K^\prime(y_2)$ and $\mbox{Re}(z)=\pm K(y_2)$ surrounding the
point
$z_2$ which now is not necessarily at $0$ as it was the case in \cite{tp89}. We
obtain $\psi_n(x)$ as a sum of two contributions of the form
\bee
\psi_n(x)=\frac{\sqrt{y_2}{\cal N}}{2\pi}\int_{-K(y_2)}^{K(y_2)}
          \left\{
           e^{xK^\prime(y_2)+ixv}
                                \frac{(t_2+i\sqrt{y_2}\mbox{sn}(v,y_2))^n}
                                     {(1+it_2\sqrt{y_2}\mbox{sn}(v,y_2))^{n+1}}
          \right. \nonumber \\
          +  e^{-xK^\prime(y_2)-ixv}  \left.
                                \frac{(t_2-i\sqrt{y_2}\mbox{sn}(v,y_2))^n}
                                     {(1-it_2\sqrt{y_2}\mbox{sn}(v,y_2))^{n+1}}
         \right\}
          \mbox{cn}(v,y_2) dv
          \nonumber\\
          +\frac{\kappa{\cal N}}{2\pi}\int_{-K(\kappa)}^{K(\kappa)}
          \left\{
           e^{ixK^\prime(\kappa)-xv}
                                \frac{(t_2\sqrt{y_2}+i\mbox{dn}(v,\kappa))^n}
{(\sqrt{y_2}+it_2\mbox{dn}(v,\kappa))^{n+1}}
          \right. \nonumber \\
         +  e^{-ixK^\prime(\kappa)+xv}  \left.
                                \frac{(t_2\sqrt{y_2}-i\mbox{dn}(v,\kappa))^n}
{(\sqrt{y_2}-it_2\mbox{dn}(v,\kappa))^{n+1}}
         \right\}
          \mbox{cn}(v,\kappa) dv
\label{eq:longpsi}
\eee
Note that in the second part of eq.\ (\ref{eq:longpsi}) the elliptic modulus
$\kappa=\sqrt{1-y_2^2}$ appears, the complementary modulus of $y_2$. As in
\cite{tp89} the second contribution in (\ref{eq:longpsi}) dominates when
$n\rightarrow\infty$ because the complex number
\[
\frac{t_2\pm i\sqrt{y_2}\mbox{sn}(v,y_2)}
     {1\pm it_2\sqrt{y_2}\mbox{sn}(v,y_2)}
\]
has a modulus smaller than $1$
\[
\sqrt{\frac{t_2^2+y_2\mbox{sn}^2(v,y_2)}
     {1+t_2^2 y_2\mbox{sn}^2(v,y_2)}}<1,
\]
since $y_2\mbox{sn}^2(v,y_2))<1$ is always fulfilled for general values of $v$.
On the contrary the modulus of the complex number
\[
\frac{t_2\sqrt{y_2}\pm i\mbox{dn}(v,\kappa)}
     {\sqrt{y_2}\pm it_2\mbox{dn}(v,\kappa)}
\]
may be larger than $1$, since $y_2\leq \mbox{dn}(v,\kappa)\leq1$, i.\ e.\
\[
\sqrt{\frac{t_2^2 y_2 + \mbox{dn}^2(v,\kappa)}
     {y_2 + t_2^2 \mbox{dn}^2(v,\kappa)}}\left\{ \begin{array}{ll}
                                                  >1 & \mbox{for $0<v<K/2$}\\
                                                  <1 & \mbox{for $K/2<v<K$}
                                                 \end{array}
                                         \right.
\]
Hence for $n\rightarrow\infty$ we only have to consider
\bee
\psi_n(x)=\frac{\kappa{\cal N}}{2\pi}\int_{-K(\kappa)}^{K(\kappa)}
          \left\{
           e^{ixK^\prime(\kappa)-xv}
                                \frac{(t_2\sqrt{y_2}+i\mbox{dn}(v,\kappa))^n}
{(\sqrt{y_2}+it_2\mbox{dn}(v,\kappa))^{n+1}}
          \right. \nonumber \\
         +  e^{-ixK^\prime(\kappa)+xv}  \left.
                                \frac{(t_2\sqrt{y_2}-i\mbox{dn}(v,\kappa))^n}
{(\sqrt{y_2}-it_2\mbox{dn}(v,\kappa))^{n+1}}
         \right\}
          \mbox{cn}(v,\kappa) dv
\label{eq:2psi}
\eee
Using that $\mbox{dn}(v,\kappa)$ and $\mbox{cn}(v,\kappa)$ are both even
functions
of the variable $v$, we derive an alternative form for the asymptotic of
$\psi_n(x)$
\be
\psi_n(x)\cong \frac{2\kappa{\cal N}}{\pi}\left\{
                                         \cos(xK^\prime(\kappa)){\cal I}_n(x) -
                                         \sin(xK^\prime(\kappa)){\cal
I}_n^\prime(x)
                                          \right\}
\ee
where ${\cal I}_n(x)$ and ${\cal I}_n^\prime(x)$ are the following integrals
\bee
{\cal I}_n(x) & = & \int_0^{K(\kappa)}\frac
  {\sqrt{y_2}\cos n\varphi+t_2\mbox{dn}(v,\kappa)\sin n\varphi}
  {y_2+t_2^2\mbox{dn}^2(v,\kappa)} \cosh(xv) \mbox{cn}(v,\kappa)\rho^n(v) dv\\
{\cal I}_n^\prime(x) & = &\int_0^{K(\kappa)}\frac
  {\sqrt{y_2}\sin n\varphi-t_2\mbox{dn}(v,\kappa)\cos n\varphi}
  {y_2+t_2^2\mbox{dn}^2(v,\kappa)} \cosh(xv) \mbox{cn}(v,\kappa)\rho^n(v) dv
\eee
with $\rho(v)$ and $\varphi(v)$ defined by
\be
\rho(v) e^{i\varphi(v)}=\frac{t_2\sqrt{y_2}+i\mbox{dn}(v,\kappa)}
                             {\sqrt{y_2}+it_2\mbox{dn}(v,\kappa)}
\label{eq:polar}
\ee
We observe that $\rho(v)$ is a decreasing function of $v$, aquiring values from
$\sqrt{\frac{t_2^2y_2+1}{y_2+t_2^2}}$
at $v=0$ to
$\sqrt{\frac{y_2+t_2^2}{t_2^2y_2+1}}$
at $v=K(\kappa)$. Since $\rho(\frac{1}{2}K)=1$ the integration intervall for
large $n\rightarrow\infty$ is practically limited to $[0,K/2]$ only and we can
approximate $\rho$ by its maximum value
$\rho\cong\sqrt{\frac{t_2^2y_2+1}{y_2+t_2^2}}$. Moreover in this interval,
$\cosh(xv)\mbox{cn}(v,\kappa)\cong1$ and peaks at a point near $v=K$. This
means
that ${\cal I}_n(x)\cong{\cal I}_n$ and ${\cal I}_n^\prime(x)\cong{\cal
I}_n^\prime$
are constants independent of $x$
\bee
{\cal I}_n & = & \left(\sqrt{\frac{t_2^2y_2+1}{y_2+t_2^2}}\right)^n
            \int_0^{\frac{1}{2}K(\kappa)}\frac
  {\sqrt{y_2}\cos n\varphi+t_2\mbox{dn}(v,\kappa)\sin n\varphi}
  {y_2+t_2^2\mbox{dn}^2(v,\kappa)} dv\\
{\cal I}_n^\prime & = &\left(\sqrt{\frac{t_2^2y_2+1}{y_2+t_2^2}}\right)^n
\int_0^{\frac{1}{2}K(\kappa)}\frac
  {\sqrt{y_2}\sin n\varphi-t_2\mbox{dn}(v,\kappa)\cos n\varphi}
  {y_2+t_2^2\mbox{dn}^2(v,\kappa)} dv
\eee
Within these approximations one may estimate the zeros of $\psi_n(x)$ in
the asymptotic limit $n\rightarrow\infty$. They are given by the equation
\be
\tan(xK^\prime(\kappa))=\frac{{\cal I}_n}{{\cal I}_n^\prime}=\tan\theta_n
\ee
or, for each $n$, the zeros $x_{np}$ are labeled by $p$
\be
x_{np}=\theta_n+p\pi
\ee
The eigenvalue problem set in eqs.\ (\ref{eq:rr1} -- \ref{eq:bc2}) is
equivalent to
\be
\tan(x_{Np}K^\prime(\kappa))=\frac{\lambda{\cal I}_{N+1}-h{\cal I}_N}
                             {\lambda{\cal I}_{N+1}^\prime-h{\cal I}_N^\prime}
                       =\tan\theta_N
\label{eq:theta}
\ee
leading to
\be
\epsilon_{Np}=\sqrt{\frac{\lambda-h^2}{y_2}}\frac{1}{K^\prime(\kappa)}
                   \{ \theta_N+p\pi\}
\label{eq:epspsi}
\ee
The spacing between the $\epsilon_{Np}$ is thus constant and there is a shift
for
each $N$ given by $\theta_N$ defined by eq.\ (\ref{eq:theta}).

A similar analysis for $\phi_n(x)$ may be given. We only quote the results
\bee
\phi_n(x)=\frac{{\cal N}}{2\pi}\int_{-K(y_2)}^{K(y_2)}
          \left\{
           e^{xK^\prime(y_2)+ixv}
                                \frac{(t_2+i\sqrt{y_2}\mbox{sn}(v,y_2))^n}
                                     {(1+it_2\sqrt{y_2}\mbox{sn}(v,y_2))^{n+1}}
          \right. \nonumber \\
          +  e^{-xK^\prime(y_2)-ixv}  \left.
                                \frac{(t_2-i\sqrt{y_2}\mbox{sn}(v,y_2))^n}
                                     {(1-it_2\sqrt{y_2}\mbox{sn}(v,y_2))^{n+1}}
         \right\}
          \mbox{dn}(v,y_2) dv
          \nonumber\\
          +\frac{i\kappa{\cal N}}{2\pi}\int_{-K(\kappa)}^{K(\kappa)}
          \left\{
           -e^{ixK^\prime(\kappa)-xv}
                                \frac{(t_2\sqrt{y_2}+i\mbox{dn}(v,\kappa))^n}
{(\sqrt{y_2}+it_2\mbox{dn}(v,\kappa))^{n+1}}
          \right. \nonumber \\
         +  e^{-ixK^\prime(\kappa)+xv}  \left.
                                \frac{(t_2\sqrt{y_2}-i\mbox{dn}(v,\kappa))^n}
{(\sqrt{y_2}-it_2\mbox{dn}(v,\kappa))^{n+1}}
         \right\}
          \mbox{sn}(v,\kappa) dv
\label{eq:longphi}
\eee
Again only the second integral dominates as $n\rightarrow\infty$ for the same
reason
as before in the expression for $\psi_n$. This last part may be recast into the
following form which incidently is proportional to $i$
\be
\phi_n(x)\cong\frac{2i\kappa{\cal N}}{\pi}\{\cos(xK^\prime(\kappa))
                                            \bar{{\cal I}}_n(x) -
                                            \sin(xK^\prime(\kappa))
                                            \bar{{\cal I}}_n^\prime(x) \}
\ee
with
\bee
\bar{{\cal I}}_n(x) & = & \int_0^{K(\kappa)}\frac
  {\sqrt{y_2}\cos n\varphi+t_2\mbox{dn}(v,\kappa)\sin n\varphi}
  {y_2+t_2^2\mbox{dn}^2(v,\kappa)} \sinh(xv) \mbox{sn}(v,\kappa)\rho^n(v) dv\\
\bar{{\cal I}}_n^\prime(x) & = &\int_0^{K(\kappa)}\frac
  {\sqrt{y_2}\sin n\varphi-t_2\mbox{dn}(v,\kappa)\cos n\varphi}
  {y_2+t_2^2\mbox{dn}^2(v,\kappa)} \sinh(xv) \mbox{sn}(v,\kappa)\rho^n(v) dv
\eee
where $\rho$ and $\varphi$ are again given by (\ref{eq:polar}). For
$n\rightarrow\infty$ the behaviour of $\rho$ limits again the integration to
the
interval $[0,K/2]$ and $\rho$ can be approximated by its maximum value
$\rho\cong\sqrt{\frac{t_2^2y_2+1}{y_2+t_2^2}}$. However here the product
$\sinh(xv)\mbox{sn}(v,\kappa)$ behaves as $xv^2$ in $[0,K/2]$ instead of beeing
nearly a constant of order unity. Thus we obtain the asymptotic behaviour for
$\phi_n(x)$
\be
\phi_n(x)\cong\frac{2i\kappa{\cal N}}{\pi}\{x\cos(xK^\prime(\kappa))
                                            \bar{{\cal I}}_n -
                                            x\sin(xK^\prime(\kappa))
                                            \bar{{\cal I}}_n^\prime \}
\ee
where now
\bee
\bar{{\cal I}}_n & = & \left(\sqrt{\frac{t_2^2y_2+1}{y_2+t_2^2}}\right)^n
   \int_0^{\frac{1}{2}K(\kappa)} v^2\frac
  {\sqrt{y_2}\cos n\varphi+t_2\mbox{dn}(v,\kappa)\sin n\varphi}
  {y_2+t_2^2\mbox{dn}^2(v,\kappa)} dv
\nonumber  \\
\bar{{\cal I}}_n^\prime & = &
\left(\sqrt{\frac{t_2^2y_2+1}{y_2+t_2^2}}\right)^n
   \int_0^{\frac{1}{2}K(\kappa)} v^2\frac
  {\sqrt{y_2}\sin n\varphi-t_2\mbox{dn}(v,\kappa)\cos n\varphi}
  {y_2+t_2^2\mbox{dn}^2(v,\kappa)} dv
\nonumber
\eee
are constants independent of $x$. The zeros $x_{np}$ of $\phi_n(x)$ are
asymptotically given by
\be
\tan(x_{np}K^\prime(\kappa))=\frac{\bar{{\cal I}}_n}{\bar{{\cal I}}_n^\prime}
                            =\tan\bar{\theta}_n
\ee
The boundary conditions (\ref{eq:bc1}) and (\ref{eq:bc2}) of the
CTM problem yield the eigenvalues
$\epsilon_{Np}$
\be
\tan(q^{-1}\epsilon_{Np}K^\prime(\kappa))=\tan\bar{\theta}_N
=\frac{\bar{{\cal I}}_{N+1}-h\bar{{\cal I}}_N}
      {\bar{{\cal I}}_{N+1}^\prime -h\bar{{\cal I}}_N^\prime}
\ee
or
\be
\epsilon_{Np}=\frac{q}{K^\prime(\kappa)}\{\bar{\theta}_N + p\pi\}
\label{eq:epsphi}
\ee
We observe again that the levels $\epsilon_{Np}$ are equidistant in this limit
of large $N$,
but there is a translation by an amount
$\frac{q}{K^\prime(\kappa)}\bar{\theta}_N$ for each polynomial of order $N$.

Eqs.\ (\ref{eq:epspsi})  and (\ref{eq:epsphi})
are the main results of this paper,
again confirming, for the simplest vertex model, the results of Baxter
\cite{baxtersbook} in an explicit calculation for a finite system.

A last remark concerns the level spacing. From (\ref{eq:epspsi}) and
(\ref{eq:epsphi}) we have
\be
\Delta\epsilon=\epsilon_{N,p+1}-\epsilon_{N,p}=\frac{q\pi}{K^\prime(\kappa)}
                                              =\frac{q\pi}{K(y_2)}
\label{eq:spacing}
\ee
The modulus used here is $y_2$ of eq.\ (\ref{eq:y2}). For $h=0$ we have
$y_2=\lambda$
which is thus different from the modulus parametrization used in \cite{tp89}
which
was $\lambda^{-1}$. Moreover the normalization of the energy levels is also
different
due to the fact that the Carlitz polynomials are directly used in \cite{tp89}.
The
Carlitz polynomials are normalized so that the first one is always $1$ or $x$
depending upon the parity. Here our $\psi_0(x)$ is not $1$ but contains an
$x$ dependence according to eq.\ (\ref{eq:psi0}) which may be divided out
later.

At $h=0$ where $y_2=\lambda$ and $t_2=0$ there is a decoupling in eqs.\
(\ref{eq:longpsi}) and (\ref{eq:longphi}), respectively, and we recover the
results
of \cite{tp89}.

As in \cite{et92} we have checked the level spacing (\ref{eq:spacing})
numerically
with standard methods (cf.\ \cite{wilkinson77}) of diagonalising the
pentadiagonal matrix which is equivalent to the recursion relations
(\ref{eq:rr1}) and (\ref{eq:rr2}) if either $\psi_n$ or $\phi_n$ is eliminated.
We observed equidistant level spacing to rather high accuracy already for very
moderate system sizes of the order of $N=20$.

\section{Summary and conclusion}
The generator of the CTM of a generalized free Fermion vertex system of finite
size is a quantum spin chain Hamiltonian with particular interactions which
increase linearly along the chain. We have
presented the analytical diagonalisation of this particular quantum spin chain
in the asymptotic regime of large system size $N$ for arbitrary values of the
parameters, the anisotropy $\lambda$ and the magnetic field $h$,
in the region where $\lambda>h^2$.

Let us briefly summarise the methods applied to accomplish our goal and restate
our main result for easy reference.
The asymptotic diagonalisation has been achieved through the explicit
construction of a new class
of elliptic polynomials which are the components of the eigenvectors of the
problem.
In this construction an elliptic parametrization of the generating functions of
the polynomials has been used which is based on the treatment of a
two--parameter
elliptic integral. The asymptotic evaluation of an integral representation of
these polynomials yields the eigenvalues, given by eqs.\ (\ref{eq:epspsi})
and (\ref{eq:epsphi}), respectively, which are equidistantly spaced with
spacing
\[
\Delta\epsilon=\frac{q\pi}{K(y_2)}
\]
the modulus $y_2$ of the complete elliptic integral $K(y_2)$ being related
to the generating functions of the eigenvectors and given explicitly in
(\ref{eq:y2}). This equidistant spacing is the main result of the present work
extending the findings of previous studies \cite{et92,tp90,tp88,tp90} to
general values of the parameters and thereby confirming once again the
general expectation \cite{baxtersbook}.

We have not touched the issue of the orthogonality of the polynomials
$\psi_n(x)$
and $\phi_n(x)$ which may be called associate Carlitz polynomials. Since the
three
limiting cases are orthogonal polynomials, it is natural to expect that the
``associated Carlitz polynomials'' remain orthogonal. Presumably the proof is
based
on the continuous fraction expansion of some elliptic functions interpolating
between the Jacobian elliptic functions $\mbox{cn}(x,k)$ and $\mbox{dn}(x,k)$
\cite{carlitz60}. We have not succeded in proving this yet.

\section*{Acknowledgements}
H-P E would like to thank the Laboratoire de Mod\`eles de Physique
Math\'ematique
of the Universit\'e de Tours for hospitality and financial support
during several stages of this
work, Professor L Turban, Drs.\ B Berche and J-M Debierre for discussions and
the Laboratoire de Physique du Solide of the Universit\'e de Nancy I
for hospitality and financial support during
the final stage of this work.

\section*{Appendix}
As we have seen the main problem has been the inversion of the elliptic
integral of
{\ref{eq:ei}}. The solution obtained was given in eq.\ (\ref{eq:tshift}) which
upon replacing $t_2$ by its expression reads
\[
t=\frac{h(1-y_2) - i(1-\lambda y_2) \sqrt{y_2}
                \mbox{sn}(iqu,y_2)}
       {(1-\lambda y_2) - i h(1-y_2) \sqrt{y_2}
                \mbox{sn}(iqu,y_2)}
\]

In this appendix we check the limit $\lambda\rightarrow1$ against a direct
calculation. Let us assume $\lambda=1-\epsilon$, then
$y_2=1-\frac{\epsilon}{\sqrt{1-h^2}}$ or using $h=\cos\theta$ as in ref.\
\cite{tp90}
$y_2=1-\epsilon/\sin\theta$. Then to first order in $\epsilon$
$(1-\lambda y_2)\cong\epsilon(1+1/\sin\theta)$, and we have
\[
\lim_{\lambda\rightarrow1}\mbox{sn}(iqu,y_2)=\tanh(iu\sin\theta)
\]
and
\[
\lim_{\lambda\rightarrow1}t=\frac{\cos\theta+(1+\sin\theta)\tan(u\sin\theta)}
                                 {(1+\sin\theta)+\cos\theta\tan(u\sin\theta)}
\]
which upon inversion gives $u$ as a function of $t$
\[
u(t)=\frac{1}{2i\sin\theta}\ln\left(\frac{(1+\sin\theta-t\cos\theta)+
                                          i(\cos\theta-(1+\sin\theta)t)}
                                         {(1+\sin\theta-t\cos\theta)-
                                          i(\cos\theta-(1+\sin\theta)t)}\right)
\]
or
\[
u(t)=\frac{1}{2i\sin\theta}\ln\left(\frac{t-e^{-i\theta}}{t-e^{i\theta}}\right)
        +\frac{\theta+\pi/2}{2\sin\theta}
\]
which agrees with the $u(t)$ directly computed from the integral (\ref{eq:ei})
with
$\lambda=1$.

Next we check the limit $\lambda\rightarrow h^2$. There we have
\[
y_2\cong\frac{\lambda-h^2}{(1-h^2)^2}\rightarrow 0
\]
and
\[
\lim_{y_2\rightarrow0}i\sqrt{y_2}\mbox{sn}(iqu,y_2)=-\frac{\sqrt{\lambda-h^2}}
{(1-h^2)}\sinh((1-h^2)u)
\]
Now to obtain the correct limit we must impose a shift
\[
u(1-h^2)=x(1-h^2)-\frac{1}{2}K^\prime(y_2)
\]
Then as $y_2\rightarrow0$
\[
\lim_{y_2\rightarrow0}=x(1-h^2)-\frac{1}{2}\ln(4/y_2)=x(1-h^2)-\ln\left(
                             \sqrt{\frac{4(1-h^2)^2}{\lambda-h^2}}
                                                                  \right)
\]
leads to
\[
\sinh((1-h^2)u)\cong-\frac{1}{2}\frac{2(1-h^2)}
{\sqrt{\lambda-h^2}}e^{-x(1-h^2)}
\]
Finally for $\lambda\rightarrow h^2$
\[
\lim_{y_2\rightarrow0}i\sqrt{y_2}\mbox{sn}(iqu,y_2)=e^{-x(1-h^2)}
\]
which is want one obtains by direct integration.

The correctness of the two limits implies that the limiting generating
functions
are generating functions for Meixner polynomials of first and second kind
according
to the classification of Chihara \cite{chihara78} or the Gottlieb and Meixner
Pollaczek polynomials according to an independent classification.

\end{document}